**Why accumulation mode organic electrochemical transistors turn off much faster than they turn on**

*Jiajie Guo,[1][†] Shinya E. Chen,[1][†] Rajiv Giridharagopal,[2] Connor G. Bischak,[2] Jonathan W. Onorato,[3] Kangrong Yan,[4] Ziqiu Shen,[4] Chang-Zhi Li,[4] Christine K. Luscombe,[3,5] David S. Ginger[2]\*.*

[1] Molecular Engineering and Science Institute, University of Washington, Seattle, WA 98195, United States.
[2] Department of Chemistry, University of Washington, Seattle, WA 98195, United States.
[3] Department of Materials Science and Engineering, University of Washington, Seattle, WA 98195, United States.
[4] State Key Laboratory of Silicon and Advanced Semiconductor Materials, Department of Polymer Science and Engineering, Zhejiang University, Hangzhou 310027, P.R. China.
[5] pi-Conjugated Polymers Unit, Okinawa Institute of Science and Technology Graduate University, Onna-son, Okinawa, 904-0495, Japan.

[†] These two authors contributed equally to the work.
**\*** Corresponding author e-mail, dginger@uw.edu

**Abstract**

Understanding the factors underpinning device switching times is crucial for the implementation of organic electrochemical transistors (OECTs) in neuromorphic computing and real-time sensing applications. Existing models of device operation cannot explain the experimental observations that turn-off times are generally much faster than turn-on times in accumulation mode OECTs. Through operando optical microscopy, we image the local doping level of the transistor channel and show that device turn-on occurs in two stages, while turn-off occurs in one stage. We attribute the faster turn-off to a combination of engineering as well as physical and chemical factors including channel geometry, differences in doping and dedoping kinetics, and the physical phenomena of carrier density-dependent mobility. We show that ion transport is limiting the device operation speed in our model devices. Our study provides insights into the kinetics of OECTs and guidelines for engineering faster OECTs.

**Main**

Organic electrochemical transistors (OECTs) are currently being explored for applications including bioelectronics,[1–4] logic circuit elements,[5,6] and neuromorphic devices.[7–10] As a class of transistors, OECTs feature high transconductance (≈ mS),[11] low operation voltage (typically < |1 V|),[11] and direct response to biologically relevant ions[12,13] and neurotransmitters.[14,15] The typically soft and flexible nature of organic semiconductors used in OECTs enables the detection of action potentials[2,16] and opens the window for applications in brain-machine interfaces and in vivo sensing.[17–19] To unleash the full potential of OECTs, a deeper understanding of the fundamental transistor operation mechanism is necessary, especially transistor switching behaviors, which are critical to the training phase of neuromorphic computing and simulating behaviors of arrays of transistors and their scaling properties.[20,21]

In OECTs, organic semiconductors — also referred to as organic mixed ionic-electronic conductors (OMIECs) in this context — are used as channel active layers, with the most common materials being conjugated polymers.[22–26] The conductivity of an OECT is modulated by the electrochemical gate potential, which controls the doping level (redox state) of the conjugated polymer channel. Importantly, in contrast to conventional field-effect transistors (FETs), OECTs exhibit volumetric doping: the gate voltage changes the conductivity of the entire volume of the transistor channel, rather than just the surface layer, and counterions injected from the electrolyte provide charge compensation for injected electronic carriers.[27] At steady-state, the channel current ($I_\text{D}$) is governed by both carrier mobility and carrier density. The steady-state behavior of OECTs has been relatively well studied.[22,26,28,29] To benchmark the device performance, the product of electronic mobility and volumetric capacitance, $\mu C^*$, has been recognized as the material figure of merit of OECT in steady-state operation.[30]

Compared to the steady-state performance, our current knowledge of OECT kinetics is limited.[29] For example, the switching speed for materials with identical $\mu C^*$ can vary by many orders of magnitude.[31] Understanding the switching behaviors of OECTs is crucial for designing logic units as well as emulating and sensing neural activity, which typically operates at the frequency of ≈ 100 Hz.[16] The widely-used Bernards model describes the transient behavior with an equivalent RC circuit as the ionic path and makes the quasistatic approximation for the channel charge distribution.[28] Several improved models based on the original Bernards model have been proposed with more complex equivalent circuits describing the ionic circuit.[32,33] Recently, Paudel et al.[34] demonstrated a 2D-finite element model based on drift-diffusion process and shows the existence of the lateral ion current during switching, which is neglected in the traditional model. However, it should be noted that all these kinetics studies are based on poly(3,4-ethylenedioxythiophene):poly(styrene sulfonate) (PEDOT:PSS), which operates as a depletion mode OECT, with the transistor in the on-state when no gate potential is applied. The transient response of PEDOT:PSS-based OECTs may not be directly comparable to accumulation mode OECTs. Accumulation-mode OECTs are generally more favorable for practical applications because of their comparatively lower energy-consumption, particularly for in vivo bioelectronics and neuromorphic computing. Nevertheless, to our best of knowledge, no studies have yet to systematically discuss the transient response and device turn-on/turn-off kinetics of accumulation mode OECTs.

In this article, we study the asymmetric transient behavior of accumulation mode OECTs. Through operando optical microscopy coupled with OECT characterization, we find that OECT turn-on occurs in two distinguishable stages: (1) doping front propagation and (2) vertical doping. In contrast, device turn-off occurs in one continuous stage. We further identify the factors that contribute to the rapid OECT turn-off behavior including typical faster dedoping kinetics of conjugated polymers compared to their doping kinetics, the channel geometry, and the carrier density-dependent mobility. Combining these observations, we propose an empirical model describing the switching behavior of accumulation mode OECTs and provide physical interpretations to the response time constants. Lastly, we show that ion transport is the limiting factor to device kinetics, and we offer guidance for engineering faster accumulation mode OECTs from both materials and device perspective.

## Asymmetric OECT response times

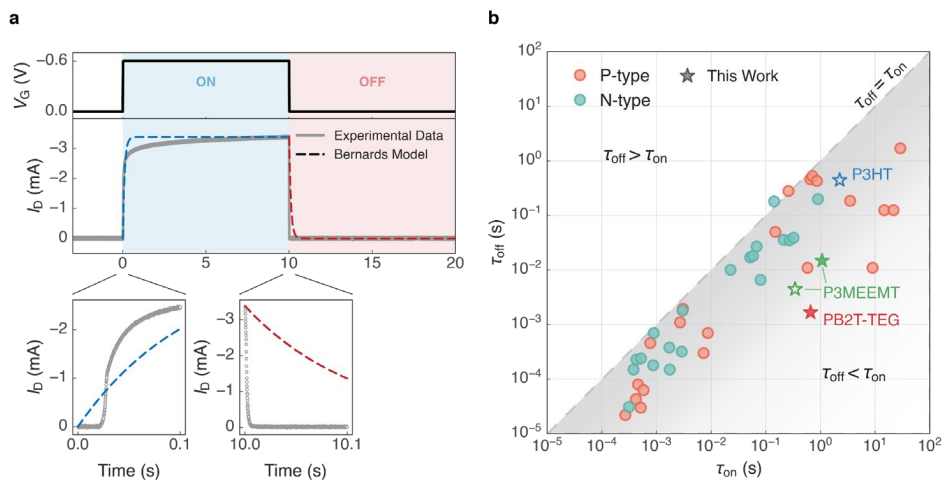

**Fig. 1 | OECT response times. a**, Transient response of a typical accumulation mode OECT (solid) and the fit with Bernards model (dashed). **b**, Accumulation mode OECT response times in literatures. Each point represents one polymer-electrolyte pair. Detailed response times and references are listed in **Table 1**. The device response times measured in this work with carefully chosen gate voltage considering threshold voltage are represented as star symbols, which are PB2T-TEG (0.1 M KCl), P3MEEMT (0.1 M KCl), P3MEEMT (0.1 M KTFSI) and P3HT (0.1 M KTFSI).

**Figure 1a** shows a typical transient response of an accumulation mode OECT: the transistor is turned on (higher $|I_D|$) upon gate potential applied and is turned off upon potential removal. We first tried the Bernards model[28] to describe the transistor switching behavior (**Supplementary Note 1**). The Bernards model expresses the variation of $I_D$ over time as a single exponential function upon a square $V_G$ pulse, with one RC time constant related to ion transport into the channel polymer.[28] We found that Bernards model fails to predict the transient response of accumulation mode OECTs from three aspects: (1) the initial transistor turn-on, which manifests as a short delay (**Fig. 1a**, bottom left) rather than instant turn-on, cannot be well-captured; (2) the transistor turn-on cannot be described with a single exponential function; (3) the difference in transistor switching-on and switching-off times cannot be described (**Fig. 1a**). These issues, especially the difference in transistor switching times, cannot be resolved even if the improved models are applied, which focus more on interpreting the pre-exponential factor.[32,33]

**Figure 1b** and **Table 1** display the turn-on and turn-off times of accumulation mode OECTs from both this paper and our literature survey. We considered published results with typical planar structure and aqueous electrolyte, which is the most common OECT structure to date. **Figure 1b** shows that faster device turn-off compared to turn-on is indeed ubiquitous, though rarely discussed aside from limited reports on PEDOT:PSS.[35]

One hypothesis is that this asymmetry could arise from the switching potentials chosen. Based on the Butler–Volmer model, the electrochemical reaction rate is influenced by the activation potential.[36–38] Since the threshold voltage ($V_T$) varies with different polymer-electrolyte systems, it is possible that the faster turn-off is the result of smaller voltage difference between $V_{on}$ and $V_T$ compared to $V_{off}$ and $V_T$, namely, $|V_{on} - V_T| < |V_{off} - V_T|$. To rule out the influence of the mismatch between switching on and off gate potentials, we selected three p-type polymers as examples and carefully tested their OECT kinetics with fixed voltage difference between $V_T$ and switching potentials. The three conjugated polymers studied here are: poly[2,5-bis-(thiophenyl)-1,4-bis(2-(2-(2-methoxyethoxy)ethoxy)ethoxy)-benzene] (PB2T-TEG), poly(3-{[2-(2-methoxyethoxy)ethoxy]methyl}-thiophene-2,5-diyl) (P3MEEMT), and poly(3-hexylthiophene-2,5-diyl) (P3HT); and the two aqueous electrolytes are: potassium chloride (KCl) and potassium trifluoromethanesulfonimide (KTFSI). The chemical structure of the polymers and the device performance are shown in **Supplementary Fig. 1**. After considering threshold voltage, we still observed faster turn-off behavior (**Fig. 1b**, stars). Clearly, the faster turn-off is not caused by the mismatch between switching potentials, and the asymmetry in OECT switching times may be related to other factors such as polymer doping and dedoping kinetics or device geometry.

## Comparison between OECT and spectroelectrochemistry

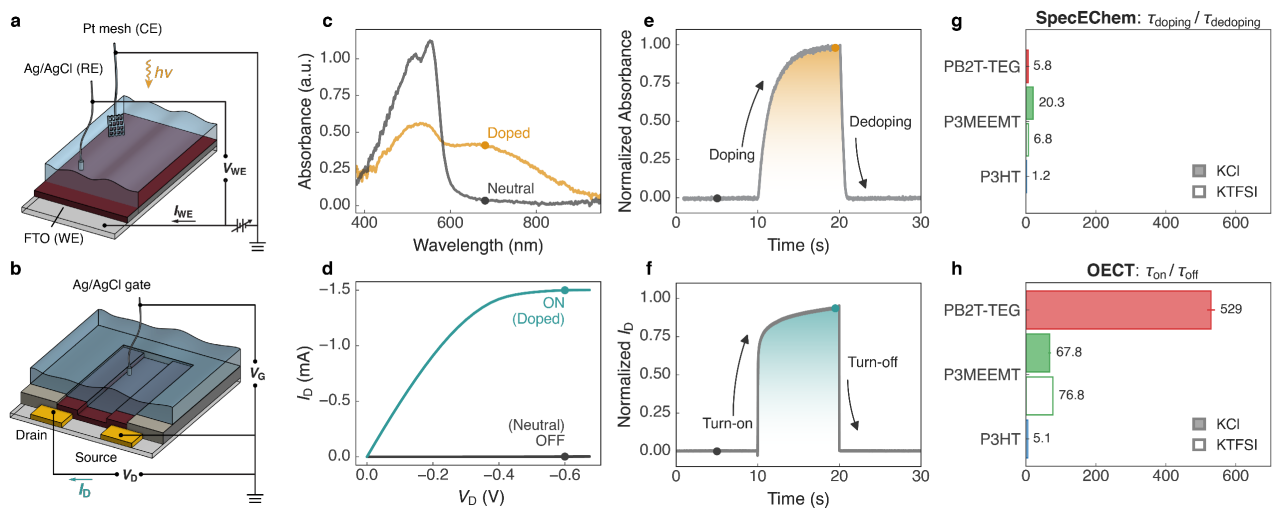

**Fig. 2 | Comparison between OECT and UV-Vis spectroelectrochemistry. a,b**, Schematic diagram of spectroelectrochemistry (SpecEChem) (a) and OECT (b) experimental setups. **c,d**, Typical UV-Vis spectra (c) and OECT output curves (d) of two steady-states (neutral/doped). **e,f**, An example transient response of spectroelectrochemistry (e) at polaron peak absorption wavelength (680 nm) and OECT (f) at saturation region ($V_D = -0.6$ V). **g,h**, The $\tau_{doping}/\tau_{dedoping}$ measured in spectroelectrochemistry (g) and the $\tau_{on}/\tau_{off}$ measured in OECT (h) for: PB2T-TEG and P3MEEMT with 0.1 M KCl (solid); P3MEEMT and P3HT with 0.1 M KTFSI (unfilled) The doping-voltage for spectroelectrochemistry, $V_{doping} = $ OCP + 0.2 V, and the dedoping-voltage, $V_{dedoping} = $ OCP − 0.2 V. The on-voltage for OECT, $V_{on} = V_T - 0.2$ V, and the off-voltage, $V_{off} = V_T + 0.2$ V. Time resolution of spectroelectrochemistry is ≈ 10 ms and OECT is ≈ 10 μs. Error bars represent standard error of the mean from multiple cycles.

Because the magnitude of $I_D$ is closely related to the electrochemical doping level of the channel conjugated polymer, it is possible that faster OECT turn-off is the consequence of rapid electrochemical dedoping nature of the polymer. To verify this hypothesis, we compared the doping and dedoping kinetics of spectroelectrochemistry (two-terminal diodes) to the turn-on and turn-off speed of OECTs (three-terminal transistors). **Figure 2a,b** display the geometries of the spectroelectrochemistry and OECT measurements. **Figure 2c** shows the steady-state UV-Vis spectra, which provides information on the electronic states of the conjugated polymers. When electrochemically doped, the polymer is oxidized along with the formation of a polaron, resulting in the bleaching of the π-π* transition peak (≈ 525 nm) and the increase of the polaron peak (≈ 680 nm). **Figure 2d** shows the transistor output curves associated with the on-state (doped) and the off-state (neutral). In addition to the steady-state study, the time-resolved UV-Vis spectra provide the rate of polaron formation and removal, which we expressed as the time constants: $\tau_{doping}$ and $\tau_{dedoping}$. We obtained the time constants from exponential fittings of polaron absorption peak over time (**Fig. 2e** and **Supplementary Figs. 2** and **3**). Below, we referred to similar time constants from the transient response of OECT switching as $\tau_{on}$ and $\tau_{off}$ in order to distinguish them from the spectroelectrochemistry time constants (**Fig. 2f** and **Supplementary Figs. 2** and **3**). Given the effect of the activation potential discussed in the previous section, we carefully measured the operando UV-Vis spectra under the same potential difference with respect to the equilibrium potential, or open circuit potential (OCP). The $V_T$ and OCP values are listed in **Supplementary Table 1**.

**Figure 2g,h** summarize the ratios of $\tau_{doping}/\tau_{dedoping}$ (spectroelectrochemistry) and $\tau_{on}/\tau_{off}$ (OECT) of four polymer-electrolyte pairs, respectively. We showed that, in both spectroelectrochemistry and OECT, the processes involving polymer doping are slower than the ones associated with polymer dedoping. However, across all polymers and electrolytes, the switching difference between these two processes is much larger in OECTs compared to spectroelectrochemistry. **Supplementary Table 2** reports the collected response times. Interestingly, we found that the timescale of OECT turn-on is comparable to the timescale for spectroelectrochemical doping, while OECT turn-off is much faster than spectroelectrochemical dedoping (approximately 10-100x faster). These results suggest that faster OECT turn-off is not simply due to faster polymer dedoping. We next turn to explore the possible causes.

## Operando microscopy characterization

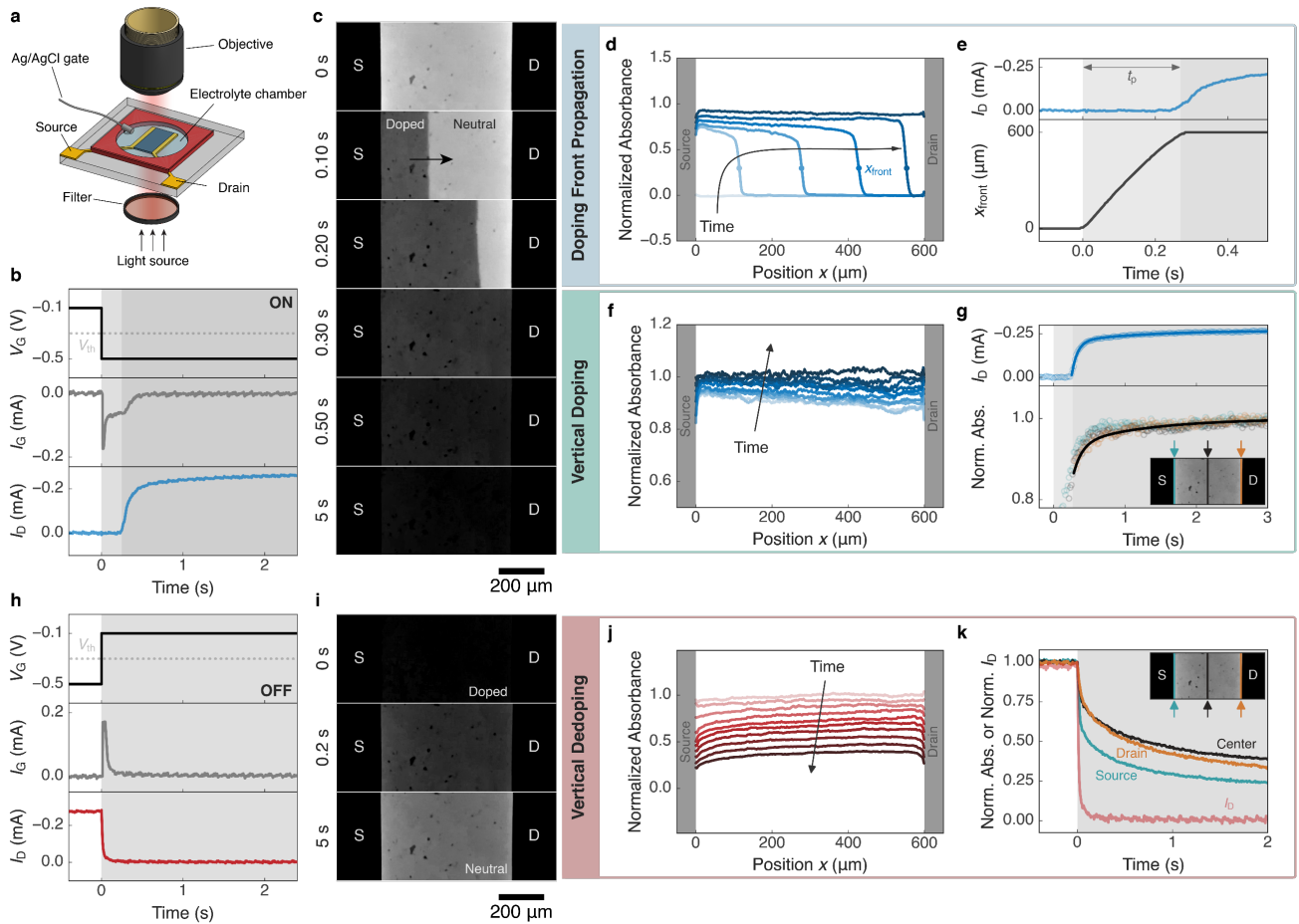

**Fig. 3 | Operando optical microscope coupled with OECT switching. a**, Operando optical microscope setup schematics. **b**, Potential and current response during turn-on. **c**, Microscope movie screenshots during turn-on with timestamp labels. Darker pixel represents higher polaron concentration. **d**, Normalized polaron absorption along channel over time during *doping front propagation* stage (stage 1 in turn-on). **e**, Comparison of $I_D$ and moving front position over time during *doping front propagation* stage. **f**, Normalized polaron absorption along channel over time during *vertical doping* stage (stage 2 in turn-on). **g**, Comparison of $I_D$ and normalized polaron absorption over time at selected positions during *vertical doping* stage. The solid lines indicate the fits. The insert image shows the selected positions over the channel. **h**, Potential and current response during turn-off. **i**, Microscope movie screenshots during turn-off with timestamp labels. **j**, Normalized polaron absorption along channel over time during turn-off, or *vertical dedoping*. **k**, Comparison of normalized $I_D$ and normalized polaron absorption at three selected positions over time. The insert image shows the selected positions over the channel (labeled as source, center and drain). A 650 nm long pass filter was used, and the red channel intensity is used to calculate the polaron absorbance. The polymer used here is PB2T-TEG, and the electrolyte is 0.1 M KCl. The transistor channel length is 600 μm, and the film thickness is ≈120 nm. The drain potential is fixed at -0.6 V.

To further understand the origin of faster device turn-off in accumulation mode OECTs, we probed the electrochemical doping level of channel conjugated polymer via operando optical microscopy coupled with a 650 nm long pass filter to selectively monitor polaron formation (**Fig. 3a**). **Figure 3b** shows the transient response of $I_D$ and $I_G$ during transistor turn-on. We observed an immediate $I_G$ response to the $V_{on}$, which shows a typical spiking and decay behavior, suggesting ion injection from the electrolyte into the channel polymer. In contrast, $I_D$ remains relatively low initially after $V_{on}$ applied, which to our knowledge, has not been discussed previously. **Figure 3c** shows the microscope movie screenshots during turn-on. Darker pixels represent more polaron absorption and thus a higher electrochemical doping level. We found that OECT device turn-on occurs in two stages: (1) a *doping front propagation* stage and (2) a *vertical doping* stage. We provide detailed discussions of each stage in the following paragraphs.

During the *doping front propagation* stage, we found that even though substantial doping of the channel polymer is already occurring, the growth of the $I_D$ starts only after the doping front position ($x_{front}$) reaches the drain electrode (**Fig. 3d,e** and **Supplementary Video 1**). We thus introduce the doping front propagation time, $t_p$, as the time required for the doping front to propagate across the entire channel from the source to the drain electrode, with the value of $\approx$ 270 ms for this particular device ($L$ = 600 μm). We defined $x_{front}$ as the position of the peak of the first derivative of the absorbance. **Figure 3e** shows the relatively linear relation between $x_{front}$ and time, suggesting the front is moving at a constant speed, which is $\approx$ 2.2 μm/ms in this case. The fact that we observed the doping front propagating from the source to drain electrode suggests that the injection of electronic carriers is occurring primarily from the source electrode during device turn-on. This result makes sense considering p-type OECTs are typically operated at negative $V_G$ and $V_D$, and electron removal (or hole injection) from channel conjugated polymer (polymer oxidation) will naturally favor the grounded source electrode instead of negatively biased drain electrode. In addition, we point out that this doping front propagation stage ($I_D$ remains relatively low after $V_{on}$ applied) is typically overlooked if characterized with standard source meters such as a Keithley 2400 ($\approx$ 250 ms resolution, see SI for discussion).

In the *vertical doping* stage, we observed a similar speed of doping of the polymer in the center of the channel and near both electrodes (**Fig. 3f,g**). We found the speed of $I_D$ increase ($\tau_{OECT} \approx$ 290 ms) and polymer doping ($\tau_{SpecE} \approx$ 200 ms) is at the same order of magnitude, suggesting the $I_D$ increment is largely dominated by the increase of doping level (or carrier density) along the channel with the underlying polymer acting as an increasingly conductive, planar electrode. **Figure 3** shows data from PB2T-TEG, a polymer that undergoes a distinct structural phase transition upon doping.[39] While the ion-induced phase transition in PB2T-TEG provides an easily resolvable doping front for kinetic analysis, we found the two-stage behavior of a doping front propagation followed by a vertical doping, is a general across accumulation mode OECT behavior regardless of cycle numbers (**Supplementary Video 2**), active layer material (**Supplementary Video 3**), or channel length (**Supplementary Fig 4**).

For the OECT device turn-off, or the *vertical dedoping* stage, we observed an immediate response of both $I_D$ and $I_G$ to the $V_{off}$ (**Fig. 3h** and **Supplementary Video 4**). We did not observe any front propagation event compared to turn-on. **Figure 3i,j** display the polymer dedoping process along the channel. While we did not see a clear dedoping front, quantitative analysis of the microscope images shows that polymer does dedope faster near the source electrode than other positions in the channel (**Fig. 3k**). This finding partly explains the faster OECT turn-off. The conductive channel is broken by the dedoping of a thin slice near the electrode: if we consider the transistor channel as a series of resistors, as long as the resistance of one of the resistors increases (the polymer near source electrode), the total resistance will increase and thus result in the decrease in $I_D$. Nevertheless, this explanation must be incomplete, as the turn-off speed of $I_D$ is *still* much faster than the dedoping speed of polymer near source electrode. As a corollary of this behavior, we found that even though the transistor is in the off-state already, dedoping of the channel polymer is not fully completed as noted by the absorbance in the polaron band (**Fig. 3k**). This phenomenon suggests that, through shortening the off-interval (duty cycle), faster OECT turn-on in the subsequent cycle can be readily achievable, as the channel polymer is already in slightly doped state (**Supplementary Note 2**). We speculate that this behavior may be useful in particular for designing OECT-based spiking neural networks[7,8] and emulating neuron dynamic filtering function.[40,41] We next considered whether carrier-density dependent mobility can account for the remaining acceleration of the turn-off time.

## Carrier density-dependent mobility

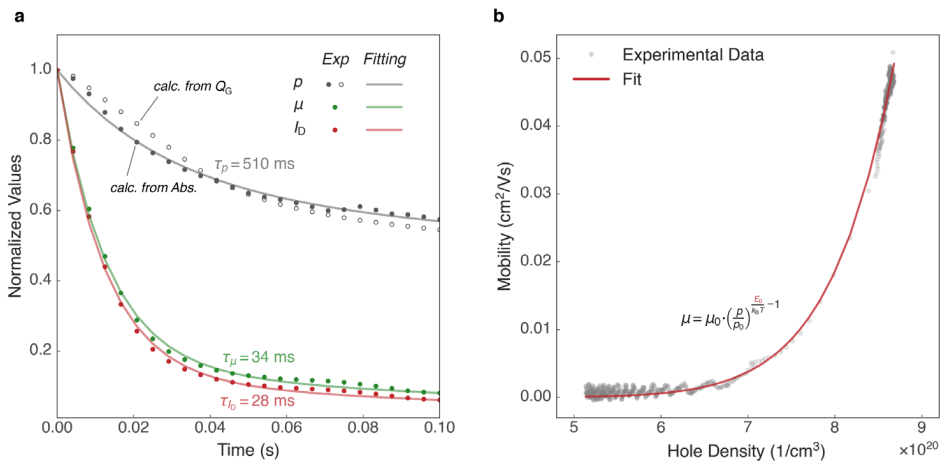

**Fig. 4 | OECT mobility and carrier density**. **a**, Transient response of $I_D$, carrier density and average carrier mobility during the PB2T-TEG transistor turn-off. Solid line indicates the fit with biexponential equation. **b**, Calculated carrier mobility as a function of carrier density using eqn (1). Solid line indicates the fit with the eqn (2). The good fit of the classic density-dependent mobility formula to the experimental data based on the measured current and carrier density suggests the importance of carrier density dependent mobility in explaining the rapid turn-off of OECTs.

**Figure 4a** shows the transient response of the $I_D$, carrier density and average carrier mobility during the transistor turn-off. We calculated the carrier density using both the integral of gate current over time and average polaron absorbance along the channel during the vertical dedoping stage and both results are in good agreement (**Supplementary Fig. 5**). We then estimated the average hole mobility in the channel during turn-off via:

$$\mu = \left(\frac{L}{W \cdot d \cdot e \cdot V_D}\right) \cdot \frac{I_D}{p} \qquad (1)$$

assuming a linear electric field along the channel, where $\mu$ is the average carrier mobility, $p$ is the hole density, $e$ is the electron charge and $V_D$ is the drain voltage. $L$, $W$, and $d$ represent channel length, width, and thickness, respectively. We found the turn-off rate of $I_D$ is comparable to the rate of the carrier mobility decay, which, due to the non-linear relationship between density and mobility in conjugated polymers, is about one order of magnitude faster than the carrier removal rate. This result agrees with what we have observed previously that the OECT turn-off speed is about an order of magnitude faster than polymer dedoping speed measured by spectroelectrochemistry (**Supplementary Table 2**).

**Figure 4b** shows the calculated carrier mobility and extracted carrier density relation required to explain our data. We found that carrier mobility is indeed carrier density-dependent, especially in the high carrier density region. Previously, Friedlein et al. had demonstrated that the steady-state OECT performance could be well characterized if carrier density-dependent mobility is considered, with the relation between mobility and density being:

$$\mu = \mu_0 \cdot \left(\frac{p}{p_0}\right)^{\frac{E_0}{k_B T} - 1} \qquad (2)$$

where $\mu_0$ is mobility prefactor and $p_0$ is zero-filed hole concentration. $E_0$ describes the energetic width of the tail of the density of states, $k_B$ is Boltzmann's constant and $T$ is temperature.[42] This equation captures the filling of the density of states (DOS) due to energetic disorder in conjugated polymer materials.[43] We found this relation also fits our carrier mobility and density data well. **Supplementary Note 3** describes detailed fittings and discussions of Eqn 2. We concluded that the carrier density-dependent mobility explains why the OECT turn-off is even faster than the dedoping of polymer near source electrode: not only carrier density, but also the carrier mobility decreases significantly at the initial stage of device turn-off, and both carrier mobility and carrier density contribute to $I_D$.

## Engineering faster OECTs

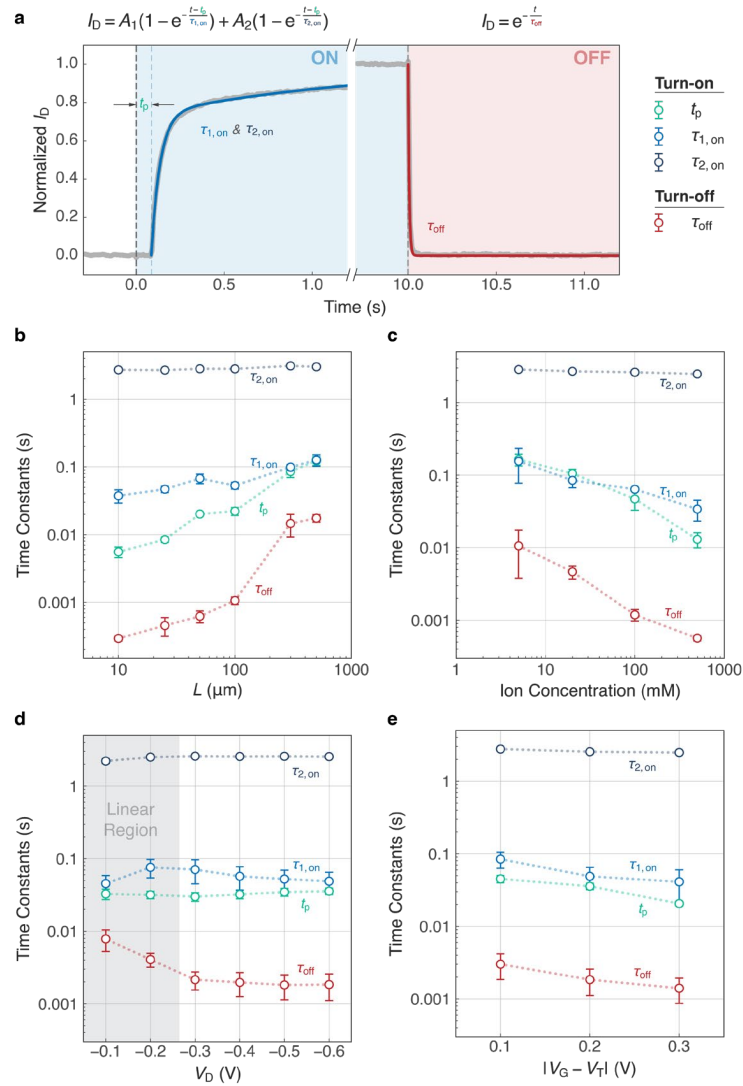

**Fig. 5 | Dependency of OECT response times on the operation variables. a**, Transient response of a typical accumulation mode PB2T-TEG OECT and the fitting equations. **b**, Relation between time constants and channel length. **c**, Relation between time constants and ion concentration. **d**, Relation between time constants and $V_D$. **e**, Relation between time constants and $|V_G - V_T|$. Transistor channel width is 2.5 or 5 mm, and the thickness is ≈50 nm for all cases. Electrolyte is 0.1 M KCl and transistor channel length is 100 μm unless otherwise specified. $V_D$ is −0.6 V and $|V_G - V_T|$ is 0.2 V unless otherwise specified. Error bars are standard error of the mean from at least 3 different devices. Dashed lines are guide to the eye.

With the transient behavior knowledge from operando microscopy, we expressed the two-stage turn-on with an empirical biexponential equation including the initial front propagation time ($t_p$):

$$\left|I_{D,\,\text{Norm}}(t)\right| = A_1 \cdot \left(1 - e^{-\frac{t-t_p}{\tau_{1,on}}}\right) + A_2 \cdot \left(1 - e^{-\frac{t-t_p}{\tau_{2,on}}}\right); \quad A_1 + A_2 = 1 \qquad (3)$$

where $\tau_{1,\,on}$ and $\tau_{2,\,on}$ are time constants associated with the vertical doping stage. $A_1$ and $A_2$ are two constants with the value of $A_1$ typically $\approx 0.7$ (**Fig. 5a**). For one-stage transistor turn-off, we described $I_D$ as:

$$|I_{D,\,Norm}(t)| = e^{-\frac{t}{\tau_{off}}} \quad (4)$$

where $\tau_{off}$ is the time constant expressing vertical dedoping (**Fig. 5a**).

To understand how the factors in Eqn. 3 depend on experimental conditions, we next studied how operation variables including potential and device geometry affect OECT switching behavior (**Fig. 5b–e** and **Supplementary Figs. 6–12**). During the *doping front propagation* stage, we found smaller $t_p$ with shorter channel length as expected because the doping front is propagating at a relative constant speed. Surprisingly, we found $t_p$ is relatively independent of the drain potential. This result suggests that ion transport from electrolyte into the polymer channel, instead of electronic transport from the source electrode, is limiting the front propagation speed. Indeed, we found faster $t_p$ values with higher gate potentials, as the increased gate potential speeds up ion movement. We further demonstrated smaller $t_p$ with increased ion concentration, thinner channel active layer and bulky anion,[44,45] all associated with shorter ion transport time from electrolyte into polymer channel.

In the *vertical doping* and *vertical dedoping* stages, we found a similar trend of $\tau_{1,\,on}$ and $\tau_{off}$ compared to $t_p$, namely, smaller $\tau_{1,\,on}$ and $\tau_{off}$ if the ion transport time from electrolyte into polymer layer is reduced (higher gate potential, increased ion concentration, thinner polymer layer and bulky anion[44,45]). We hypothesize that smaller $\tau_{1,\,on}$ and $\tau_{off}$ with shorter channel length is akin to charging and discharging a capacitor, where smaller capacitance (shorter channel length) results in faster charging and discharging. In contrast, the slower $\tau_{2,\,on}$, with the magnitude of $\approx$ 2-3 s, is less dependent on all operation variables. We propose that $\tau_{2,\,on}$ is associated with polymer structural relaxation or ionic/electronic charge reorganization, as recently suggested by Wu et al.[46] In addition, we noticed that when $|V_G - V_T| > |V_D|$, or when the device is operated in the linear region (shadowed area in **Fig. 5d** and **Supplementary Fig. 13**), the difference

between turn-on and turn-off response time becomes smaller. This result is reasonable as the transistor behaves like a diode-like spectroelectrochemistry device when drain potential is decreased.

Finally, we demonstrated a SPICE circuit model that accurately reflects transient asymmetry of accumulation mode OECTs by incorporating a time-dependent channel resistor into the existing model.[33,47] The detailed SPICE simulation methods and results are in **Supplementary Note 4**. To sum up, from the device perspective, shorter channel length, thinner polymer layer and higher gate potential (not drain potential) facilitate more rapid device switching. From the materials perspective, one could increase ion concentration, select a bulky counter anion, or design polymer with high ionic conducting ability or with rigid backbone.[48,49]

**Conclusions**

We show that three different conjugated polymers exhibit much faster turn-off compared to turn-on when used to make accumulation mode OECTs. This behavior, while ubiquitous in the literature, is rarely discussed, and is inconsistent with many common OECT models. Using operando optical microscopy, we demonstrate that device turn-on occurs in two temporally and spatially distinct stages: first, a doping front propagates from the source to the drain; second, the partially doped channel continues to dope more homogeneously. In contrast, turn-off occurs in a single step, with the kinetics varying weakly across the channel and the fastest dedoping occurring near the source. We identify several factors contributing to faster device turn-off including channel geometry, differences in doping and dedoping kinetics, and the physical phenomena of carrier density-dependent mobility. In this limit, reducing channel length can be very important to reducing turn-on time due to the initial doping front propagation step, however this might sacrifice the transconductance of OECTs. Second, doping processes are generally slower than dedoping, a factor we hypothesize is due to the rearrangement of the polymer that accompanies the doping process. Nevertheless, the doping front propagation and asymmetry of doping and dedoping rates are still insufficient to explain the large differences in turn-on vs. turn-off times for OECTs. The final key piece which we propose to explain the rapid turn-off is density-dependent mobility. Notably, the functional form of the mobility that we extract using this hypothesis is in excellent agreement with previous reports of the function dependence of carrier mobility on charge density in organic semiconductors. Finally, we show that ion transport is limiting the device speed in that it controls both doping and dedoping kinetics. These observations provide guidelines for engineering faster accumulation mode OECTs from both materials and device perspectives. We believe this unique transient asymmetry renders accumulation mode OECT as a tailorable "slow-learning, fast-forgetting" unit which may found position in either the toolbox of traditional circuit design or neuromorphic computing applications. We anticipate these results will aid in the selection of counterion chemistries and transistor geometries for specific applications and set that stage for full drift-diffusion models that simulate device behavior of both accumulation and depletion mode OECTs, and improve the accuracy of simplified equivalent circuit models.

**Methods**

**Polymer film preparation and characterization**

The synthesis of PB2T-TEG[39,50] (detailed discussion on molecular weight in the original paper[50]) and P3MEEMT[45] ($M_n$ = 24 kg/mol, Đ = 1.67) polymers were described in our previous works. P3HT was obtained from Ossila (M109, $M_w$ =36.6 kg/mol). PB2T-TEG polymer was dissolved in chloroform with the concentration of 2 to 4.5 mg/mL. P3MEEMT and P3HT polymers were dissolved in chlorobenzene with the concentration of 20 mg/mL. All polymer solutions were stirred overnight at 50 °C prior to spin coating. The substrates were cleaned sequentially by sonication in acetone and isopropanol for 15 min each. The surface of the substrate was then treated with oxygen plasma for 3 min before spin coating. The spin rate used is 600–2500 rpm to control film thickness between 20 nm to 120 nm. PB2T-TEG polymer films were annealed at 150 °C for 10 minutes under $N_2$ after spin-coating.

**OECT device fabrication and characterization**

OECT devices comprised lithographically patterned gold on glass substrates (see lithography process below) with transistor widths of 2.5 mm, 5 mm or 6 mm and lengths ranging from 10 μm to 600 μm. Polymers were spun casted onto OECT substrates and were carefully removed except at the electrode junction region via cotton tips (slightly dampened with acetone solution) under microscope or magnifying lens to ensure minimum impact on the transient response. A secure seal hybridization chamber (GRACE BIO LABS) is attached onto the substrate to confine the electrolyte. A Ag/AgCl pellet is used as the gate. The distance between gate and channel is fixed at approximately 4 mm in this study. The transfer curves were measured using two Keithley 2400 source-measure units controlled by custom Python code. The transient measurements were conducted with NI PXIe-5451, NI PXIe-6366 and NI PXIe-8381 controlled by custom LabVIEW code with time resolution ≈ 10 μs.

The detailed lithography process: NR9-3000PY negative resist (Futurrex, Inc.) was deposited on cleaned glass wafers with diameter equals to 100 mm (University Wafer, Inc.) using Rite Track Automated

Coater (SVG-90S), followed by UV light exposure (ABM-SemiAuto-Aligner) and resist development (SVG-90S). Metal deposition (10 nm chromium and 100 nm gold) was accomplished through evaporation (CHA Solution e-beam evaporator). The resist lift-off was achieved by soaking wafers in acetone solution overnight. The wafers were then diced using a Disco Wafer Dicer (Disco, America).

**Operando microscope coupled with OECT characterization**

An iPhone 11 Pro (1080 P, 240 fps) was attached to Leica CME microscope with a 15× eyepiece, a 10× objective (NA = 0.25) and a 650 nm long pass filter (FEL0650, THORLABS) for video recording. Videos were analyzed using custom Python code with OpenCV library. A dark and a reference image were taken for dark and flat field frame corrections, and absorbance calculation. To optimize the video contrast, a thicker PB2T-TEG film was prepared by drop casting from 1 mg/mL chlorobenzene solution to slow down the evaporation rate.

**Spectroelectrochemistry characterization**

The ultraviolet-visible (UV-Vis) absorption spectra were measured using an AVANTES spectrometer (AvaSpec-2048L) coupled with an AVANTES light source (Avalight-HAL-S). Doping and dedoping UV-Vis spectra were collected using continuous mode (with AvaSoft software) with time resolution ≈10 ms/spectrum. The potential bias is controlled using a Metrohm Autolab PGSTAT204 (with NOVA Software version 2.1). Polymers were cast onto fluoride-doped tin oxide-coated glass (FTO, Sigma-Aldrich, 7 Ω/sq) and used as a working electrode. A Ag/AgCl electrode and a Pt mesh were used as reference electrode and counter electrode, respectively. All three electrodes were submerged into a cuvette containing ≈2.5 mL of either 0.1 M $KCl_{(aq)}$ or 0.1 M $KTFSI_{(aq)}$.

**Data availability**

The data that support the findings of this study are available from the corresponding author on request.

**Acknowledgements**

This paper is based on research supported primarily by the National Science Foundation, DMR-2003456. K.Y., Z.S., and C.-Z.L. thank the support from National Natural Science Foundation of China (22125901) for supporting synthesis of the PB2T-TEG polymer. J.W.O. and C.K.L.'s contributions to P3MEEMT polymer synthesis are based in part on work supported by the National Science Foundation DMREF-1922259. Part of this work (transistor fabrication) was conducted at the Washington Nanofabrication Facility/Molecular Analysis Facility, a National Nanotechnology Coordinated Infra- structure (NNCI) site at the University of Washington with partial support from the National Science Foundation via awards NNCI-1542101 and NNCI-2025489.

**Author contributions**

J.G. and S.E.C. contributed equally to the work. J.G., S.E.C. and D.S.G. conceived the project, designed the experiments, and discussed the results together. J.G. and S.E.C. performed the experiments and analyzed the data. S.E.C. wrote the first draft and J.G. made the figures. R.G. performed the SPICE circuit modeling. C.G.B designed the preliminary microscope experiment. K.Y., Z.S., and C.-Z.L. provided the PB2T-TEG polymer. J.W.O. and C.K.L. provided the P3MEEMT polymer. J.G., S.E.C., R.G. and D.S.G. revised the manuscript with input from all the authors.

**Competing interests**

The authors declare no competing interests.

**Additional information**

**Supplementary information**

The online version contains supplementary material available at https://doi.org/xxxxxxx

**Table 1. Accumulation mode OECT response times in literatures.**

| Polymer | Type | Ion Concentration (mM) | Ion Type | $\tau_{on}$ (ms) | $\tau_{off}$ (ms) | Reference | Note |
|---|---|---|---|---|---|---|---|
| gDPP-g2T | p | PBS buffer | PBS buffer | 0.269 | 0.022 | 5 | cOECT |
| p(g2T-TT) | p | 100 | NaCl | 0.42 | 0.043 | 51 | |
| P(gTDPPT) | p | 100 | NaCl | 0.46 | 0.08 | 52 | |
| P(bgDPP-MeOT2) | p | 100 | NaCl | 0.516 | 0.03 | 53 | |
| P(lgDPP-MeOT2) | p | 100 | NaCl | 0.578 | 0.063 | 53 | |
| p(gPyDPP-MeOT2) | p | 100 | NaCl | 0.77 | 0.46 | 54 | |
| PBBTL | p | 100 | NaCl | 2.7 | 1.1 | 55 | |
| PBBTL/BBL blend | p | 100 | NaCl | 3.05 | 1.95 | 55 | |
| TDPP-gTVT | p | 100 | NaCl | 7.3 | 0.3 | 56 | |
| TDPP-gTBTT | p | 100 | NaCl | 8.7 | 0.7 | 56 | |
| PProDOT-DPP | p | 100 | LiCl | 150 | 50 | 57 | |
| PProDOT-DPP | p | 100 | $LiPF_6$ | 260 | 280 | 57 | |
| P3APPT | p | 100 | $KPF_6$ | 580 | 11 | 44 | estimation |
| PIBET-AO | p | 50 | $KCl + CaCl_2$ | 654 | 463 | 58 | |
| PIBET-O | p | 50 | $KCl + CaCl_2$ | 714 | 526 | 58 | |
| PIBET-BO | p | 50 | $KCl + CaCl_2$ | 862 | 429 | 58 | |
| PIBT-BO | p | 50 | $KCl + CaCl_2$ | 3500 | 185 | 58 | |
| P3APPT | p | 100 | KCl | 9000 | 11 | 44 | estimation |
| DPP-DTT (8:2) | p | 100 | $KPF_6$ | 14705 | 125 | 59 | 400 μm, estimation |
| DPP-DTT (1:0) | p | 100 | $KPF_6$ | 21739 | 125 | 59 | 400 μm, estimation |
| PIBET-A | p | 50 | $KCl + CaCl_2$ | 29000 | 1700 | 58 | |
| Polymer | Type | Ion Concentration (mM) | Ion Type | $\tau_{on}$ (ms) | $\tau_{off}$ (ms) | Reference | Note |
| Homo-gDPP | n | PBS buffer | PBS buffer | 0.313 | 0.031 | 5 | cOECT |
| $BBL_{152}$ | n | 100 | NaCl | 0.38 | 0.15 | 60 | |
| $BBL_{98}$ | n | 100 | NaCl | 0.43 | 0.23 | 60 | |
| $BBL_{60}$ | n | 100 | NaCl | 0.52 | 0.24 | 60 | |
| gNDI-T | n | 100 | KCl | 0.87 | 0.18 | 61 | normalized to area |
| $BBL_{15}$ | n | 100 | NaCl | 0.89 | 0.7 | 60 | |
| PBBTL/BBL blend | n | 100 | NaCl | 1.72 | 0.38 | 55 | |
| P(gTDPP2FT) | n | 100 | NaCl | 1.75 | 0.15 | 52 | |
| gNDI-V | n | 100 | KCl | 2.9 | 0.32 | 61 | normalized to area |
| P(gPzDPP-CT2) | n | 100 | NaCl | 3 | 1.8 | 62 | |
| P(gPzDPP-2T) | n | 100 | NaCl | 22.7 | 10.1 | 62 | |
| f-BTI2g-TVTCN | n | 100 | NaCl | 52 | 17 | 63 | |
| gAIID-2FT | n | 100 | NaCl | 58.5 | 18.2 | 64 | |
| f-BTI2g-TVT | n | 100 | NaCl | 68 | 27 | 63 | |
| $BBL_H$ | n | 100 | KCl | 80.3 | 6.6 | 48 | |
| $BBL_L$ | n | 100 | KCl | 142 | 181 | 48 | |
| gAIID-T | n | 100 | NaCl | 213.3 | 35.6 | 64 | |

| f-BTI2TEG-FT | n | 100 | NaCl | 272 | 35 | [65] | |
| f-BTI2TEG-T | n | 100 | NaCl | 322 | 39 | [65] | |
| BBL | n | 100 | NaCl | 900 | 200 | [66] | estimation |